\documentclass[sigplan,10pt]{acmart}

\acmConference{}
\acmYear{}

\acmISBN{} 
\acmDOI{} 
\startPage{1}

\setcopyright{none}

\usepackage{booktabs}   
\usepackage{subcaption} 
\usepackage{array}
\usepackage{amsmath}
\usepackage{url}

\usepackage{graphicx} 

\usepackage[inkscapelatex=false]{svg}

\begin{document}

\title[DSP-MLIR]{DSP-MLIR: A MLIR Dialect for Digital Signal Processing}  


\author{Abhinav Kumar}
\orcid{nnnn-nnnn-nnnn-nnnn}             
\affiliation{
  \position{Student}
  \department{School of Computing and Augmented Intelligence}             
  \institution{Arizona State University }           
  \city{Tempe}
  \state{AZ}
  \postcode{85281}
  \country{USA}                   
}
\email{akuma294@asu.edu}          

\author{Atharva Khedkar}
\orcid{nnnn-nnnn-nnnn-nnnn}             
\affiliation{
  \position{Student}
  \department{School of Computing and Augmented Intelligence}             
  \institution{Arizona State University }           
  \city{Tempe}
  \state{AZ}
  \postcode{85281}
  \country{USA}                   
}\email{atharva.khedkar@asu.edu}  

\author{Aviral Shrivastava}
\orcid{nnnn-nnnn-nnnn-nnnn}
\affiliation{
  \position{Professor}
  \department{School of Computing and Augmented Intelligence}             
  \institution{Arizona State University }           
  \city{Tempe}
  \state{AZ}
  \postcode{85281}
  \country{USA}                 
}
\email{aviral.shrivastava@asu.edu}         

\begin{abstract}
\textbf{Traditional Digital Signal Processing ( DSP ) compilers work at low level ( C-level / assembly level ) and hence lose much of the optimization opportunities present at high-level ( domain-level ). The emerging multi-level compiler infrastructure MLIR ( Multi-level Intermediate Representation ) allows to specify optimizations at higher level. In this paper, we utilize MLIR framework to introduce a DSP Dialect and perform domain-specific optimizations at dialect -level ( high-level ) and show the usefulness of these optimizations on sample DSP apps. In particular, we develop a compiler for DSP and a DSL (Domain Specific Language) to ease the development of apps. We show the performance improvement in execution time for these sample apps by upto 10x which would have been difficult if the IR were at C/ affine level.}
\end{abstract}



\maketitle

\section{Introduction}
Digital Signal Processing(DSP) is a critical technology for signal processing which is useful in many areas like audio processing( recording , mixing, denoising), speech processing (speech recognition, speaker verification, language identification) , image processing( image enhancement, compression), digital communication ( software-defined radio , radar ) to medical applications ( electrocardiogram-ECG , digital X-rays) \cite{enwiki:1233991896, dspbook}. For implementation of software DSP blocks, there is a compiler required to convert this algorithm written in high-level language into low-level programming language(i.e. object code or machine code) \cite{Aho_Lan_Sethi_Ullman_2007}. \\

There are various compiler frameworks like GCC (GNU Compiler Collection) \cite{gnu} and LLVM (Low-level virtual machine) \cite{lattner2004llvm} which provide tools to develop compilers. These traditional compiler frameworks suffer from losing the high-level abstractions as their IR (Gimple \cite{Gimple} for GCC, LLVM IR for LLVM) is more closely related to low-level language. MLIR (Multi-Level Intermediate Representation) \cite{lattner2020mlir}, an emerging compiler infrastructure under the LLVM umbrella, overcomes this by having multiple levels of IR from close to the source language down to machine-level IR. This abstraction is quite useful in specifying domain-specific computations at the desired level of abstraction and hence utilized in various domains like Quantum Computing (Quantum MLIR) \cite{quantummlir}, machine learning models (Onnx MLIR \cite{onxxmlir}, torch-mlir \cite{torchmlir}), and hardware description languages (Circt \cite{CIRCT}). Since these different dialects fall under the same infrastructure, the MLIR infrastructure seems quite promising for integrating different domains as the dialects will share consistent IR structure. This may also lead to unprecedented cross-domain optimizations as well \cite{martinez2022hdnn}.

Many modern applications combine digital signal processing (DSP) with other computational tasks, such as deep learning ie, two different domains. For example, speech recognition systems like Google Assistant, Siri, and Amazon Echo require audio preprocessing and deep learning. In medical imaging, filtering MRI scans and using deep learning for tumor detection \cite{rasool2024brain} is crucial. However, traditional compilers struggle to optimize these diverse computational tasks efficiently. MLIR's modularity allows different dialects to coexist and interact seamlessly through its unique and powerful IR structure. Currently, MLIR already includes dialects for deep learning tasks, such as onnx-mlir \cite{onxxmlir}, torch-mlir \cite{torchmlir}, and TPU-MLIR \cite{hu2022tpu}. The availability of a dedicated DSP dialect in MLIR would enable seamless integration and unlock unprecedented cross-domain optimizations, which traditional compilers cannot achieve. As a first step, this would require developing a dialect in MLIR for DSP and we make a foundational step towards this by providing a DSP Dialect in MLIR.

This paper makes the following contributions:
\begin{itemize}
    \item We present a novel MLIR Dialect for DSP and implement lowering routines to existing built-in MLIR dialects. The dialect cover operations covering a wide range of functionalities commonly used in DSP applications. Whenever possible, in order to make use of the powerful optimizations present in the Affine dialect, we lower the DSP operations to the Affine dialect. When not possible, we lower to SCF.
    \item We introduce DSP-Dialect-specific optimizations. All of the optimizations are standard well known theorems/properties in the DSP domain. Although these optimizations are straightforward to implement in the DSP dialect, but other compiler passes were not able to do them.
    \item We also provide a DSL (Domain-Specific Language) for writing DSP applications. Even though simple, this allows programmers to express their DSP applications in a programming language style (rather than writing in mlir IR format). The programs written in this DSL are converted to our MLIR DSP dialect, and then optimized and lowered from there. 
\end{itemize}

By utilizing DSP-Domain Specific properties/theorems at our dialect level, we were able to achieve upto 10x performance improvement which would have difficult for traditional compilers to achieve (ie, at C-level) . Ultimately, we deliver a comprehensive compiler for DSP in MLIR, with a frontend language to develop DSP apps using supported DSP operations and a set of dialect-specific optimizations which would automatically optimize the code if one of the patterns matches giving better performing code. This approach not only enhances the performance of DSP applications but lays down a foundation for future cross-domain integrations.
\\
The rest of the paper is divided into following sections:
Section \ref{sec:background} presents the background about MLIR and previous work for DSP compiler and MLIR dialects. In Section \ref{sec:DSPMLIRFramework}, we introduce our framework which can be categorized into 4 components -  Sec \ref{sec:DSPDialect} present our DSP dialect and  its operations; Sec \ref{sec:DSPOptimizations} present various patterns corresponding to DSP theorems/properties ; Sec \ref{sec:DSPLowering} present our approach for lowering the operations from our dialect to Affine/Scf levels and final component - Sec \ref{sec:DSPLangg} presents a DSL(Domain-Specific Language) to develop DSP applications. In Section \ref{sec:EvaluationSetup}, we present our experimental setup and evaluation strategy . In Sec \ref{sec:Results}, we present our results for various sample DSP apps. Finally, we conclude our paper by discussing future work in sec \ref{sec:Conclusion}.
\begin{figure}[ht]
    \centering
    \includegraphics[width=1\linewidth]{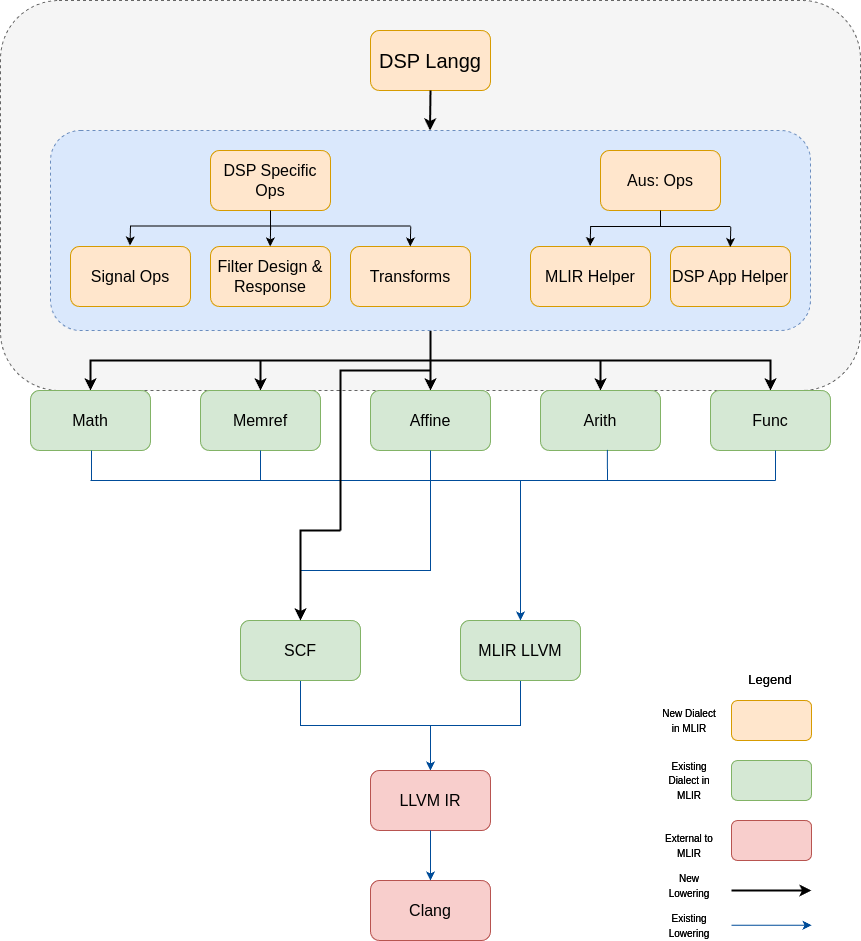}
    \caption{The compilation pipeline of our DSP language and dialect. The DSP language is first compiled into the DSP dialect, and then lowered to Affine (whenever possible) or SCF, and then to LLVM IR and Clang.}
    \label{fig:FigDSPMLIR}
\end{figure}
\section{Background and Related Work} \label{sec:background}
MLIR, an emerging compiler infrastructure and a project under LLVM umbrella introduces multi-level Intermediate Representation and its IR structure is powerful enough to represent Graph IR (Tensorflow) , Affine (C-like) to lower levels IRs like LLVM IR.  The mechanisms to engage with MLIR infrastructure are Dialects and there are dialects for different levels like tf for Tensorflow, onnx-mlir for onnx , affine for polyhedral optimizations, nvgpu for Nvidia GPUs. Dialects act as namespace for defining operations (to define functionality), types (to define datatype) and attributes (to get compile time information ex-constant data about operation) and in this work, we will introduce a new dialect for DSP domain and define set of operations required for signal processing applications.

There has been previous work in designing DSL for DSP like Fieldspar \cite{axelsson2010feldspar} (which provides dataflow style of algorithm description based on Haskell and the backend compiler produces C code and the optimizations (variable elimination, loop unrolling) are done at C-level) and FAUST \cite{letz2018faust} , another functional programming based language developed for audio processing for the Web.\\
There has been work targetting backend DSP hardware and C/assembly-level software like DSP Processors co-design with compiler \cite{DSPCompilerCodesign} and  for generating optimized code for DSP Processors using SIMD \cite{DSPSIMDOps} , Parallelization of C programs \cite{franke2003compiler}, assembly-level optimizer \cite{de2000code}, DSP processors address optimization \cite{leventhal2005dsp}, enabling auto vectorized code generation \cite{thomas2024automatic}, etc while our work targets the frontend for DSP. Matlab , a popular framework for DSP Applications also has embedded coder \cite{elrajoubi2017tms320f28335} which automatically generates fast C code for embedded processors but is closed-source.

 There have been recent works to utilize MLIR for various domains like Quantum Computing (Quantum MLIR) \cite{quantummlir}, machine learning models (Onnx MLIR \cite{onxxmlir}, torch-mlir \cite{torchmlir}), and hardware description languages (Circt \cite{CIRCT}) through specific dialects. Our work showcases another domain-specific (DSP) computations utilizing the MLIR framework. 
 
\section{Proposed DSP MLIR Framework} \label{sec:DSPMLIRFramework}
DSP Framework consists of following components - A Dialect , a DSL , Lowering and Optimizations. The compilation pipeline for our framework can be seen as Fig \ref{fig:FigDSPMLIR}.

\subsection{The DSP Dialect} \label{sec:DSPDialect}
DSP Dialect provides two set of operations - one for DSP specific blocks [like signal operations (like delay), transforms (like dft , dct , idft), filter operations (like low-pass filter design, response)] and other for auxiliary operations (like sin, cos, vector of given size generation and print) for development of DSP applications as shown in Fig \ref{fig:FigDSPMLIR} . DSP applications here represent a set of various DSP operations combined to realize applications like audio compression, low-pass noise filter etc. Sample list of Operations and Applications are provided in Table \ref{tab:dsp_ops}. MLIR already supports tensor types which is sufficient to represent DSP datatypes which are mostly one-dimensional (we will demonstrate our work on one-dimensional applications although we support multi-dimensional inherently because of tensor type which may be required for image-processing etc.) hence we use tensor type directly. 

\begin{figure*}[htbp]
    \centering
    \includegraphics[width=\linewidth]{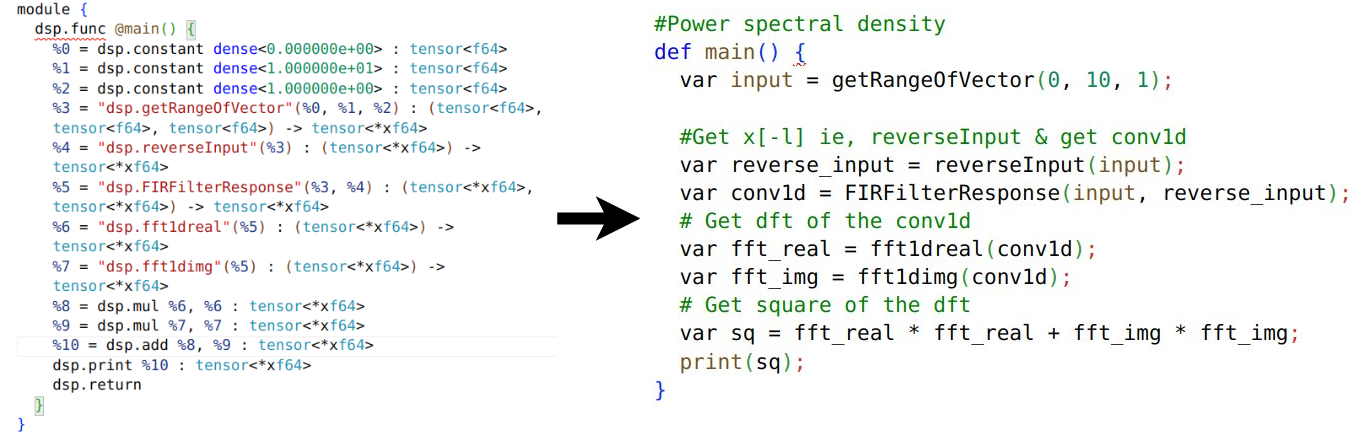}
    \caption{DSP Dialect[left] and Corresponding DSL[right] }
    \label{fig:DSPDialect}
\end{figure*}

\begin{table*}[h!]
    \centering
    \resizebox{\textwidth}{!}{%
    \begin{tabular}{|p{1cm}|p{3cm}|p{5cm}|p{6cm}|p{6cm}|}
        \hline
        \textbf{SN} & \textbf{OpName} & \textbf{Syntax} & \textbf{Equation} & \textbf{Description} \\ \hline
        1 & Delay & \tt delay(input, n) -> output & {\begin{align*}
            y[n] = x[n - k]
        \end{align*}} & Delays the input signal by \(k\) samples. \\ \hline
       2 &  FIRFilterResponse & \tt FIRFilterResponse(input, coeffs) -> output & $y[n] = \sum_{i=0}^{M} h[i] \cdot x[n-i]$ & Applies a Finite Impulse Response filter to the input signal. \\ \hline
       3 & SlidingWindowAvg & \tt slidingWindowAvg(input, window) -> output & $y[n] = \frac{1}{N} \sum_{i=0}^{N-1} x[n-i]$ & Computes the average of the input signal over a sliding window of size \(N\). \\ \hline
       4 & FFT1DReal & \tt fft1dreal(input) -> real & $X_{\text{real}}[k] = \sum_{n=0}^{N-1} x[n] \cos\left(\frac{2\pi}{N}kn\right)$ & Computes the real part of the 1D Fast Fourier Transform of the input signal. \\ \hline
       5 & FFT1DImg & \tt fft1dimg(input) -> imag & $X_{\text{imag}}[k] = -\sum_{n=0}^{N-1} x[n] \sin\left(\frac{2\pi}{N}kn\right)$ & Computes the imaginary part of the 1D Fast Fourier Transform of the input signal. \\ \hline
       6 &  LowPassFIRFilter & \texttt{lowPassFIRFilter(input, coeffs) -> output} & 
        {\begin{align*}{l}
        y_{\text{lpf}}[n] = \frac{w_c}{\pi} \cdot \text{sinc}\left(w_c \left(n - \frac{N-1}{2}\right)\right), & \text{for } n \neq \frac{N-1}{2} \\
        y_{\text{lpf}}[n] = \frac{w_c}{\pi}, & \text{for } n = \frac{N-1}{2}
        \end{align*} }
        & Applies a low-pass FIR filter to the input signal using the given coefficients. \\ \hline
        7 &  LMSFilter & \texttt{lmsFilter(input, desired, mu) -> output} & 
        {\begin{align*}
            &e[n] = d[n] - y[n] \\
            &y[n] = \sum_{i=0}^{M-1} w_i[n] x[n-i] \\
            &w_i[n+1] = w_i[n] + \mu e[n] x[n-i]
        \end{align*}} & Adaptive filtering using the Least Mean Squares algorithm. 
        \\ \hline
\end{tabular}}
\caption{Sample List of DSP Dialect Operations}
\label{tab:dsp_ops}
\end{table*}

\subsection{Domain Specific Patterns -Dialect Optimizations } \label{sec:DSPOptimizations}
\begin{figure}[ht]
    \centering
    \includegraphics[width=1\linewidth]{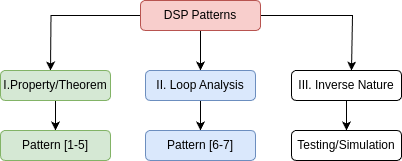}
    \caption{3 Categorizations of DSP Patterns:  Category I are the ones that are useful properties used in real world apps. Category II are loop analysis ones which aren't currently available with MLIR framework. Category III are the ones not useful in real world apps but useful for simulation purposes.}
    \label{fig:PatternCategorizatn1}
\end{figure} 
In this section, we introduce the optimizations/patterns that we do at our dialect level. MLIR provides pass infrastructure to perform transformation or optimization, and the pass infrastructure includes the pass manager, PassManager, which is responsible for organizing and executing passes at the different levels like dialect, module and function. There is also a specific kind of transformation aimed at simplifying operations (operation being the main unit of abstraction and transformation) called Operation Canonicalization in MLIR, which derives from the base pass manager. In this work, we utilize operation canonicalization to write our patterns/optimizations. 

There are two mechanisms for defining canonicalizations on operations - Rewrite Patterns and Fold. Fold mechanism is powerful mechanism and powerful at simplying operations at constant input or known properties but don't allow new operation creation. Rewrite pattern in MLIR allows for different types of canonicalizations as well as new operation creation so we use this for defining our dialect specific patterns. 

Patterns (Optimizations) in DSP dialect can be defined as replacing a group of computationally expensive dsp operations with cheaper operations. These patterns can be categorized into three categorizes as shown in Fig. \ref{fig:PatternCategorizatn1}. The first two categorizes are useful in real world applications while the third category is only useful for simulation/testing purposes and we use them to test our operations as well. The first category directly comes from DSP-Domain theorems and properties and we will discuss it in details in the next section. The second category patterns are those which are currently not supported by affine optimizations and can be done by affine transformations in the future by complex analysis, but we do these quite easily at our dialect level. We will now dive in details of each of these optimizations :

\begin{table*}[htbp]
\centering
\begin{tabular}{|m{2em}|m{7em}|m{6em}|m{15em}|c| }
    \hline
    \textbf{Opt} & \textbf{Pattern Name} & \textbf{Base OP} & \textbf{OldPattern} & \textbf{NewPattern}  \\ 
    \hline
    1 & Symmetric Filter & MulOp & [IdealFilter, Hamming, Mul] & [FilterHammOpt] \\ \hline

    2 & Symmetric Filter Response & FilterRespOp & [FilterHammOpt , FilterResponse] & [FilterResSymmOpt] \\ \hline

    3 & Filter Response at Input and Reverse & FilterRespOp & [ReverseInput , FilterResponse ] & [FilterYSymmOpt] \\ \hline

    4 & DFT Response at Symmetric Input & DFT1DRealOp & [FilterResponse , DFT1DReal ] & [DFT1DRealSymmOp] \\ \hline

    5 & Parsevaal's Theorem & DivOp & [DFT1D , Square , Sum , Div ] & [Square , Sum] \\ \hline

    6 & DFTReal and DFTImg Fusion & DFTImgOp & [DFTReal , DFTImg] & [DFT1D] \\ \hline

    7 & LMSFilter and Gain Fusion & GainOp & [LMSFilter , Gain ] & [LMSFilterGainOpt] \\ \hline
\end{tabular}
\caption{Pattern corresponding to DSP Optimizations: Here, we provide canonicalization pattern on \textit{BaseOp} and match \textit{OldPattern} and replace it with \textit{NewPattern} . Example: For row 1 , when the operands for Mul is IdealFilter and HammingWindow , we replace the 3 operations (shown in col - \textit{OldPattern}) with single operation FilterHammOpt (shown in col - \textit{NewPattern}). }
\label{tab:dsp_pattern}
\end{table*}

\subsubsection{Optimization 1 - Ideal filter and Cosine-Window Multiplication:} 

When designing a window-based filter, a common method is to multiply ideal low pass filter \eqref{eq:idealLowPass} with cosine-window \eqref{eq:hammingwindow} to obtain desired filter \eqref{eq:filterDesign1}. As both of them are symmetric about mid-point, the multiplication will also be symmetric and hence we have the opportunity to reduce the number of calculations to half by just calculating the first as shown in the below equation \eqref{eq:filterDesign}. Rest half is just same as first half i.e., $h[n]= h[L-1-n] $.

\begin{align}
    y_{\text{lpf}}[n] = 
            \begin{cases} 
            \frac{\omega_c}{\pi} \cdot \text{sinc}\left(\omega_c \left(n - \frac{L-1}{2}\right)\right), & \text{if } n \neq \frac{L-1}{2} \\
            \frac{\omega_c}{\pi}, & \text{if } n = \frac{L-1}{2}
            \end{cases} \label{eq:idealLowPass}
\end{align}
\begin{align}
 ham[n] = 0.54 - 0.46 \cdot cos(\frac{2\pi n}{L-1}) , 0 \leq n < L    \label{eq:hammingwindow}
\end{align}
\begin{align}
    h[n] = y_{lpf}[n] * ham[n] , 0 \leq n < L \label{eq:filterDesign1}
\end{align}
\begin{align}
    h[n] = h[L-1-n] = y_{lpf}[n] * ham[n] , 0 \leq n < \frac{L+1}{2} \label{eq:filterDesign} 
\end{align}
    
\subsubsection{Optimization 2 - FilterResponse at Noisy signal and Symmetric filter:}
Filter response operation can be defined as \eqref{eq:conv_sum1}. When the filter is symmetric, ie, $h[i] = h[L-1-i] $ then we use the symmetry property on \eqref{eq:optConv1d} to combine the beginning and end terms to obtain final \eqref{eq:optconv4}. By using \eqref{eq:optconv4} , we reduce the number of load instructions for the filter and we get performance benefits.
\begin{align}
    y[n] &= \sum_{i=0}^{L-1} 
            h[i] \cdot x[n-i] , 0 \leq n < N  \label{eq:conv_sum1}
            \\
         &= h[0] \cdot x[n] + h[1] \cdot x[n-1] + \ldots + \notag \\ 
         &\quad h[L-2] \cdot x[n-(L-2)] + h[L-1] \cdot x[n-(L-1)] \\
         &= h[0] \{ x[n] + x[n-(L-1)] \} +  \notag \\ 
         &\quad h[1] \{ x[n-1] + x[n-(L-2)] \} + \ldots  \notag \\ 
         &\quad + h[\frac{L-1}{2}] \cdot x[n-\frac{L-1}{2}] \label{eq:optConv1d} \\
         &= \sum_{i=0}^{\frac{L-1}{2}} h[i] \cdot \{ x[n-i] + x[n-(L-1-i)] \} \notag \\ 
         &\quad +  h[\frac{L-1}{2}] \cdot x[n-\frac{L-1}{2}] \label{eq:optconv4}
\end{align}


\subsubsection{Optimization 3 - FilterResponse/Conv1D Property at input and reverse input:} 

According to filter response property, we know that when the inputs are vector and reverse vector ie, $h[l] = x[-l] or, h[l] = x[L-1-l] $ then the output of filter response will be symmetric about its mid-point ie, $\frac{L+1}{2} $ so we check the pattern for the operands of filter response and if the inputs are  reverse of each other, we calculate the output for first half only as shown in the below equation \eqref{eq:conv_symm} and for the second half, we use $ y[n] = y[N-1-n] $. Hence we achieve performance benefit by reducing the number of loop iterations. 

\begin{align}
    y[n] &= \sum_{i=0}^{L-1} 
            h[i] \cdot x[n-i] , 0 \leq n < N  \label{eq:conv_sum} 
\end{align}

\begin{align}
    y[n] = y[N-1-n] = \sum_{i=0}^{L-1} h[i] \cdot x[n-i] , 0 \leq n < \frac{N+1}{2}  \label{eq:conv_symm}  
\end{align}

\subsubsection{Optimization 4 - DFT Response at Symmetric Input:}

According to DFT Property , when the input is real and symmetric , then the real part of DFT will be symmetric and the imaginary part will be conjugate symmetric as shown in \eqref{eq:DftReal} and \eqref{eq:DftImg} so this also gives an opportunity to reduce the outer loops to half. Here, the symmetry happens after the first element.
\begin{align}
    X[k] &= \sum_{n=0}^{N-1} x[n] e^{-j \frac{2\pi}{N} nk}, \quad k = 0, 1, 2, \ldots, N-1 \\
    &= X_{real}[k] + j X_{img}[k]
\end{align}
\begin{equation}
    X_{real}[k] = X_{real}[N-k], \quad k = 1, 2, \ldots, \frac{N-1}{2} \label{eq:DftReal}
\end{equation}
\begin{equation}
    X_{img}[k] = -X_{img}[N-k], \quad k = 1, 2, \ldots, \frac{N-1}{2} \label{eq:DftImg}
\end{equation}

\subsubsection{Optimization 5 - Parsevaal's Theorem:} 

According to Parseeval's theorem as shown in \eqref{eq:ParsevaalTh}, energy of a signal is equal in time as well as frequency domain. So, whenever there is a pattern in which energy is calculated in frequency domain, we can replace the same with that in time domain which means if we find a code in which there is sum of square of real and imag part of DFT is calculated ( as shown in Table \ref{tab:dsp_pattern} (Opt 5) ) , we replace it with sum of square of the input.
\begin{equation}
    \sum_{n=0}^{N-1} |x[n]|^2 = \frac{1}{N} \sum_{k=0}^{N-1} |X[k]|^2 \label{eq:ParsevaalTh}
\end{equation}

\subsubsection{Optimization 6 - Loop fusion for DFTReal and DFTImg Part:} 

Currently, with our DSL just supports returning a single result so we calculate DFTReal and DFTImg part separately according to \eqref{eq:dftRealCos} and \eqref{eq:dftImgSin} respectively. When the inputs to both the operations are same, affine loop fusion should fuse the operations into one but we observe it is unable to do so. Hence, we write our own pattern to fuse these operations into one , overall saving loop iterations. 
\begin{align}
    X_{\text{real}}[k] &= \sum_{n=0}^{N-1} x[n] \cos\left( \frac{2\pi}{N} k n \right), 0 \leq n < N \label{eq:dftRealCos}\\
    X_{\text{img}}[k] &= -\sum_{n=0}^{N-1} x[n] \sin\left( \frac{2\pi}{N} k n \right) , 0 \leq n < N \label{eq:dftImgSin}\\
    X[k] &= X_{\text{real}}[k] + j X_{\text{img}}[k] \
\end{align}

\subsubsection{Optimization 7 - LMSFilter and Gain: }

LMS Filter is an adaptive filter that aims to minimize the mean square error between the desired signal \( d(n) \) and the output signal \( y(n) \).
The filter is widely used for applications such as hearing aid, noise cancelling and echo cancelling among others.
LMS Filter can be represented as the following equations:

1. \textbf{Output Signal:}
\begin{equation}
y(n) = \mathbf{w}^T(n) \mathbf{x}(n)
\end{equation}
where
\begin{itemize}
    \item \( \mathbf{w}(n) = [w_0(n), w_1(n), \ldots, w_{M-1}(n)]^T \) is the weight vector at time \( n \).
    \item \( \mathbf{x}(n) = [x(n), x(n-1), \ldots, x(n-M+1)]^T \) is the input vector at time \( n \).
    \item \( M \) is the number of filter taps.
\end{itemize}

2. \textbf{Error Signal:}
\begin{equation}
e(n) = d(n) - y(n)
\end{equation}
where
\begin{itemize}
    \item \( d(n) \) is the desired signal at time \( n \).
\end{itemize}

3. \textbf{Weight Update:}
\begin{equation}
\mathbf{w}(n+1) = \mathbf{w}(n) + \mu e(n) \mathbf{x}(n)
\end{equation}
where
\begin{itemize}
    \item \( \mu \) is the step size (learning rate) parameter.
\end{itemize}

As seen from the above equations, the calculation of LMSFilter weights is a two-dimensional nested loop  while gain operation is a one-dimensional loop computation. LMS and Gain blocks are used together for hearing aid application where the gain can be fused with LMSFilter as follows: 

Weight Update with Gain:
\begin{equation}
\mathbf{w}(n+1) = \mathbf{w}(n) + \mu G e(n) \mathbf{x}(n)
\end{equation}
where \( \mu \) is the step size (learning rate) parameter and \( G \) is the gain applied to the error signal.

However traditional compilers such as affine optimization fails to see this opportunity which we exploit in our dialect. \\

Table \ref{tab:dsp_pattern} shows the patterns with dsp operations corresponding to each of these optimizations and the corresponding new pattern.

Here, first 5 patterns are the category one patterns directly coming from DSP Domain property/theorem. Pattern 6 and 7 are category two patterns. The category three patterns are the patterns like upsampling followed by downsampling , transforms followed by its inverse(like {fft, ifft} , {dct, idct} ) which are mostly used for testing and simulation of dsp operations but would have really difficult to do at C or lower-levels. 
\\

\subsection{The DSP Dialect Lowering}
\label{sec:DSPLowering}

Once the MLIR compatible IR is obtained, the next step is to define the lowering of operations defined in DSP-Dialect, which means the actual implementation of operations. For DSP operations, the implementation will require loop-based computations hence the choices in MLIR Framework are either affine dialect or scf dialect. Affine dialect in MLIR promises powerful polyhedral optimizations like loop fusion, ScalarReplacement hence we choose this dialect for lowering. But this dialect has certain restrictions like it requires constant value for loop-indices and symbols which is not applicable in certain scenarios like dynamic index value calculation, then the lowering is done to scf loop directly. For implementation, we need to have understanding of affine IR structure and syntax and the corresponding affine C++ API for our framework development. MLIR framework is still in its early phase and is emerging and the C++ APIs are evolving and hence the development and testing can be quite slow for the developers. For example, there are multiple ways to implement same functionality through C++ API for ex- there are api's for implementing for loop in affine like \textit{affine::buildAffineLoopNest} and \textit{affine::AffineForOp}. The second one \textit{affine::AffineForOp} is more powerful in expressing affine loops and which one to choose needs well-documented programmer's guide which is currently lacking from MLIR side. Similarly, for binding affine references, one may either use \textit{bindDims()} or the references can be obtained directly by \textit{PatternRewriter.getAffineDimExpr()} and which one to use requires proper documentation from framework developers. Mlir is also evolving and certain feature usage like returning multiple value types from control block needs more complex analysis ie, \textit{scf::IfOp} block using \textit{TypeRange\{floatType , indexType\}} and we need to reiterate affine-IR for another simpler code and this can lead to longer development time. Overall, the development of mlir C++ code is harder, hence we need an approach for our lowering so that the development is smoother. 

\begin{figure}[ht]
    \includegraphics[width=0.8\linewidth]{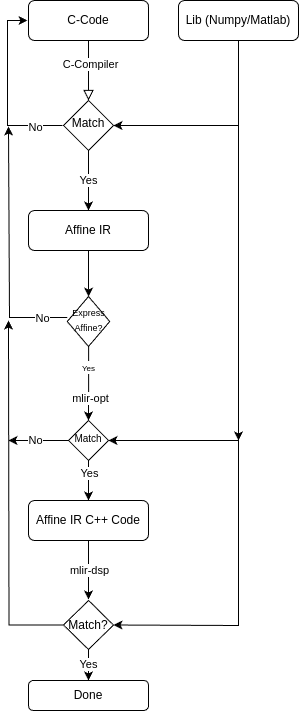}
    \caption{Approach for Developing Mlir C++ Lowering : There are 3 stages- C-code (easiest), Affine IR and finally C++ code (hardest). We develop C-code , match against std library , then develop Affine IR , and if IR can't be expressed with C-code, go back to developing C-code. Once IR output matches with std library, we move to final stage of C++ code development until output matches. }
    \label{fig:AffineLowering}
\end{figure}

We developed an approach using multiple stages to make our development smoother as shown in Fig \ref{fig:AffineLowering} where the stages are C-level, Affine-level IR and affine C++ code. The complexity level increases as we go down the stages hence we need to validate each stage so that our implementation is functionally correct. For each stage, we develop the code and compare the output with standard libraries like numpy/Matlab. The validation at each stage also makes our debugging easier which gets harder as we go to lower stages. The first stage is C-level code which is quite easy to develop and can be easily compiled \& tested with corresponding dsp libraries. The second stage is Affine-level IR and it is C-like but with slight differences and restrictions so if our C-code can't be expressed as affine loop iterations , we again move to stage first and try another C-code resolving those restrictions like SSA restrictions, constant index etc. We try C-code iteration till we can express our c-code in terms of Affine-IR then this Affine-level IR can be tested with mlir-opt which is available from MLIR framework and validated against desired output. Once the output of this stage matches with that of standard libraries, we can go for our final C++ development which is hardest (as C++ API is very dynamic and can change a lot) but this is needed for our framework development. In this stage, we check the existing api (through corresponding C++ definition ) and get the corresponding usage ( through MLIR source code APIs and other existing lowering ) and develop the proper C++ code and use \textit{mlir->dump()} for matching against second-stage output. This final stage output can be obtained by emitting affine-IR from the our framework and we match against the output of std library  (or, the second stage Affine-IR not shown in the fig \ref{fig:AffineLowering} for simplicity) . 

Once the lowering is done, the rest of the lowering is done using built-in MLIR lowering passes. MLIR framework provides lowering from affine/scf loops to mlir-llvm (a dialect in MLIR). We use this and then there is a conversion pass which translates the mlir-llvm IR to LLVM IR. Once this LLVM IR is obtained , clang tool is used to convert this LLVM IR into executable as shown in the fig 4. This executable is run and validated against desired output. 

\subsection{The DSP Domain-Specific Language} \label{sec:DSPLangg}
Using the lowering and clang-17, DSP applications can be developed and tested but in MLIR, dialects have to be written at MLIR IR level and is not user-friendly as shown in Fig \ref{fig:DSPDialect} and hence affects productivity from a programming perspective. Hence, we introduce a simple user interface language, DSP DSL to ease the development. The grammar for DSP DSL Language consists of simple syntax rules [like statement ending with semicolon] and tokens [like keywords(def , main, var), arithmetic operators(+,-,*,/), braces , square bracket for tensors ]. 
An example application for calculating power spectral density of a signal is shown in Fig \ref{fig:DSPDialect} - right part. As demonstrated in the Fig \ref{fig:DSPDialect}, this code is simple but robust enough to ease the programming and testing of DSP dialect operations. 
The conversion steps are usual as any standard compiler stages :a) Lexer produces the token/symbols which are then used by recursive Parser to generate AST (Abstract Syntax Tree). Parser parses the module (source file) , which is made of functions and function is made of statements and produces moduleAST. Then, there is another Parser (mlirGen class) which takes the moduleAST as input and generates corresponding operation based on statement types from DSL AST which will generate the final MLIR IR as an output, which can be easily fed into MLIR system for further processing/analysis. 
\section{Evaluation Setup} \label{sec:EvaluationSetup}
\begin{figure*}[htbp]
    \centering
    \includegraphics[width=\linewidth]{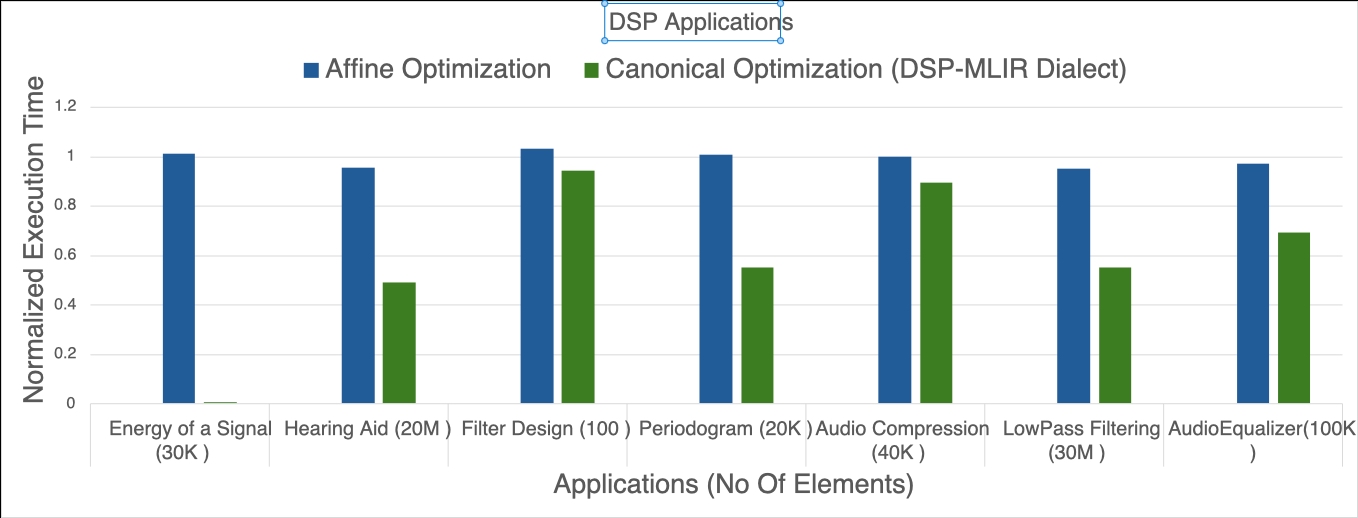}
    \caption{Normalized Performance with canonical optimizations in DSP-MLIR}
    \label{fig:results}
\end{figure*}

\begin{table*}[htbp]
\centering
\begin{tabular}{|m{1em}|m{7em}|m{20em}|m{9em}|m{2em}| }
    \hline
    \textbf{SN} & \textbf{AppName} &  \textbf{DSP Ops Sequence} & \textbf{Optimizations} & \textbf{Opt} \\ 
    \hline
    1 & Filter Design & [Ideal LowPass Filter -> Hamming -> Mul] & [Symmetric Filter] & [1] \\ \hline

    2 & LowPass Filtering & [Input + Noise -> Filter Design -> Filter Response ] & [Symmetric Filter and FilterResponse] & [1,2] \\ \hline

    3 & Energy of Signal & [Input -> DFT -> Square -> Sum ] & [Parsevaal's Theorem] & [5] \\ \hline

    4 & Spectral Analysis & [Input , Reverse -> Conv1D -> DFT -> Square ] & [Filter Response and  DFT Symmetric] & [3,4] \\ \hline

    5 & Audio Compression & [Input -> DFT -> Threshold -> Quantization -> RunLenEncoding ] & [DFT Loop Fusion] & [6] \\ \hline

    6 & Hearing Aid & [Input , Ideal LMS -> LMSFilter -> Gain -> LMSFilter Response ] & [LMSFilter and Gain Fusion] & [7] \\ \hline

    7 & Audio Equalizer & [Input , LowPass , BandPass , HighPass -> Gain for Bands -> FilterResponse -> Sum] & [Symmetric Filter and FilterResponse] & [1,2] \\ \hline
    
\end{tabular}
\caption{ Sample DSP Apps and sequence of the DSP operations in the app and the optimizations applied to them. }
\label{tab:DspApps}
\end{table*}

We evaluate DSP-MLIR framework on sample DSP Apps (here, an application is input, set of DSP blocks/operations and output ) which are representative of Audio signal processing but can be easily applied to other digital signal processing as well. 

The test applications were executed on a machine having AMD Ryzen 5700u , 8 physical cores, 16 Logical Cores, 16 GB RAM. Code version of LLVM base was 19.0.0git and version of clang was 17.0.6. 
We used two criterias for the evaluation - i) Functional Correctness, ii) Execution Time Measurement. For the first criteria ie, functional correctness, we developed similar application in numpy and since it had graphical support, we plotted the graph to make sure the desired output was as expected (ie, not only the output numbers matching but the actual application producing the desired output For example - the output signal having no ripples for low-pass filtering application ). Wherever the plots weren't applicable, we matched with the corresponding numpy output. This helped in verifying the correct sequence of the dsp operations (and the individual operations were also tested), and hence correctness of the application.  \\
For the second criteria of measuring the execution time , we had to compare our optimizations against C/affine-level optimizations ( C/affine-level  is where traditional compilers provide most optimizations) for which we enabled affine-level optimizations. For measuring the execution time of our optimizations, we also used the following methods to obtain our time - a) Took average of 5 iterations to take out system effect, b) Flushed out cache after any run so that we see same effect for every run , c) we printed out single element at random index (not the whole vector which would be an expensive operation and would increase the total time of application dramatically). 
 
\section{Results} \label{sec:Results}

In the evaluation, we ran three versions of the code - 1) without any optimization, 2) with affine optimization (AffineLoopFusion and AffineScalarReplacement) and 3) lastly with affine plus our dsp-specific optimizations for all the test apps. Table \ref{tab:DspApps} lists all the test apps and the optimizations applied on them. Fig \ref{fig:results} shows the normalized execution time versus the input size for all the applications. Here, the baseline is No Optimization (execution time = 1) and the blue and green bars are affine and affine + dsp  specific optimizations respectively.

\subsection{Our DSP lowerings and Optimizations yield much better-performing code}

As evident from the figure \ref{fig:results}, utilizing the dsp domain specific properties/theorems are much easier at dsp dialect level which yields much better performing code. This is because we are able to exploit the dsp theorem level optimizations at our dialect level which is not possible at any other lowering, performing these canonical optimizations yields around 2x performance improvement as compared to Affine optimization. 
\subsection{DSP DSL reduces programming complexity with respect to C-code}
\begin{figure}[ht]
    \centering
    \includegraphics[width=1\linewidth]{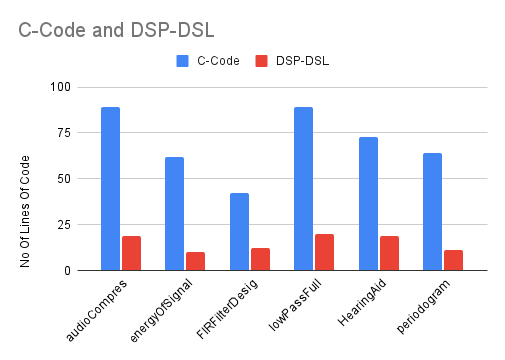}
    \caption{Lines of Code C-Code vs Our DSL for Sample Apps}
    \label{fig:CCodevsDSL}
\end{figure}



By utilizing the DSP dialect, it is easier for a signal processing engineer or compiler developer to develop and utilize DSP operations either by directly using the DSP dialect or the DSL provided within this framework. As seen from \ref{fig:CCodevsDSL}, while it may take an average of 3.5x more lines of code to write such DSP applications in C language, it is much easier and faster to implement DSP applications in our DSL. 





\subsection{DSP Domain-Specific optimizations are easier to perform at domain (high-level) }
One of the major contributions of this research is the high-level  optimizations that exploit DSP theorems and patterns to compile efficiently. These optimizations are best implemented at much higher abstraction, like a DSP dialect in MLIR. For example, for implementing the symmetric filter property at C-level in Table \ref{tab:dsp_pattern} ( Opt 1 ), this would require complex polyhedral (mathematical abstractions) analysis which would lead to complex code implementations while we do this simply by utilizing the domain knowledge about the filters. Even for loop fusion Table \ref{tab:dsp_pattern} ( Opt 6 ) , we were able to do this quite easily while affine dialect - \textit{AffineLoopFusion} and \textit{AffineScalarReplacement} optimizations were not able to fuse this double nested loops.


\section{Conclusion and Future Work}\label{sec:Conclusion}
In this work, we have developed a Domain specific compiler for DSP in MLIR and utilized the multi-level IR property to specify domain specific optimizations which was difficult to represent at lower-level IR (C/assembly). We ran these optimizations on some sample applications and obtained significant performance improvement using these optimizations with respect to affine-level optimizations.\\
\textbf{Future Work}: Existing DSP Compilers like Matlab, Scilab have graphical interfaces (like graph, 3D plots) which is quite useful in signal processing domain. Our compiler being modular can be easily integrated as the backend for these by just providing the corresponding language AST to MLIR conversion. Another interesting work will be integrating with Deep Learning blocks through available dialects like torch-mlir and look for inter-dialect optimizations and also apply in real-world applications.

\bibliography{bibfile}



\end{document}